\documentclass[12pt]{iopart}
\pdfoutput=1
\usepackage{iopams}
\usepackage{graphicx}

\begin{document}

\title[R\"uchardt revisited]{The R\"uchardt experiment revisited: using simple theory, accurate measurement and  python based data analysis.} 

\author{Chris Shearwood and Peter A. Sloan}

\address{Department of Physics, University of Bath, Bath, BA2 7AY, United Kingdom.}
\ead{p.sloan@bath.ac.uk}
\vspace{10pt}
\begin{indented}
	\item[]\today
\end{indented}

\begin{abstract}
This project uses the  R\"uchardt experiment to determine the ratio of specific heats and hence the number of degrees of freedom $f$ of different gases by measuring the frequency of damped simple harmonic motion where the gas provides the Hooke's law like spring of a cylinder-piston system. This project links mechanics, electromagnetism, thermodynamics, statistical mechanics and quantum mechanics making it an excellent synoptic experiment for a mid year undergraduate student. We present simple derivations of the main relationships that govern the experiment, a detailed data analysis of the physics of the apparatus and of the experimental data. We find 
$f(\textrm{He}) = 3.48 \pm 0.14$, $f(\textrm{N}_2) = 4.92 \pm 0.24$, $f(\textrm{Air}) = 4.96 \pm 0.25$, $f(\textrm{CO}_2) = 6.46 \pm 0.39$ at room temperature and atmospheric pressure. The results for CO$_2$ requires a statistical analysis of its vibrational modes. These results show that the expected results can be measured using fairly simple apparatus, coupled with careful analysis of large data sets.
\end{abstract}

%
\submitto{Phys. Educ.}

\section{Introduction}

\subsection{Teaching aims}
The R\"uchardt experiment links quantum mechanics, Newtonian mechanics, statistical physics, thermodynamics, and due to our set-up electromagnetism. It is an ideal mid-years undergraduate physics experiment allowing students to use the wide variety of their taught physics, experimental techniques and computational analysis they have already mastered (or at least passed assessment on). This version of implementing and analysing the R\"uchardt experiment therefore aims to bridge the gap between the more scripted and well-defined experiment they typically meet in their early years laboratory courses and the typically more open-ended, less defined projects, students tackle in their latter and final years.

Here we present an exemplar lab-report of our implementation of the R\"uchardt experiment with particular emphasis on understanding the experimental apparatus and how it transforms the physical parameter in question, to a measurable quantity. We present detailed experimental results and explore the range over which they are valid before presenting the final results. We highlight places were simple student written computer simulation would underpin and strength the student understanding of the underlying physics.  Critical to the learning success of this project as a teaching activity is that the students are already familiar to the physics concepts required for the project. It is, instead, the realistic combination of, to the students, seemingly disparate aspects of their course into a coherent scientific narrative that is the ultimate learning outcome of this project.  We  aim to publish more details, including the team based aspect of our implementation and the specifically designed authentic assessments, of this more realistic and holistic approach. 

\subsection{The R\"uchardt  experiment}

The central element of the R\"uchardt experiment is using a gas as the spring in a piston-cylinder apparatus to produce simple-harmonic-motion (SHM) of the piston when displaced from equilibrium. The frequency of that SHM can then be related to the adiabatic ratio, and hence the quantum mechanics of the gas particles.

From its initial reporting in 1929 \cite{Ruchardt1929} there has been a steady set of experimental \cite{Clark1940, Hafner1964, Smith1979, Severn2001} and theoretical investigations \cite{Gruber2004} into this simple experiment. There is a propensity to update the analysis and data-capture are more sophisticated data-capture and computer techniques become more common place in undergraduate teaching laboratories. For example on the advent of digital computing in teaching laboratories \cite{Torzo2001}.  It is available to purchase through the usual laboratory supply companies for example, LabTrek or LD Didactic. We have used it as both a year 1 and year 2 physics experiment here at the University of Bath. In the recent COVID instigated pivot to online laboratory courses it was one of the experiments to tuned into `online' experiment both at our home institutions and at Uppsala University in Sweden \cite{Matthias2021}. 

\section{Theory}

 The apparatus is lightly damped so we actually measure damped SHM (dSHM) and use the regular mathematical solutions for this system.

\subsection{Damped SHM}

From any introductory course we know that the solution to dSHM is
\begin{equation}
	z(t) = z_0 \sin (\omega_d t - \alpha) \exp\left( -t/\tau \right) \label{damped-SHM}
\end{equation}
where $z_0$ is the starting position away from the equilibrium position, $t$ is the time, $\alpha$ is some initial $t=0$ phase offset. The piston release is usually taken as $t=0$ so in general $\alpha = 0$. Finally $\tau$ is the lifetime, the time it take for damping to reduce the amplitude to $1/e$ of the initial amplitude. The damped angular frequency $\omega_d$ is related in the normal way to the undamped angular frequency $\omega_0$ by 
\begin{equation}\label{damped_correction}
	\omega_d = \sqrt{\omega_o^2 - \left(1/\tau \right)^{2}}
\end{equation}
with
\begin{equation}
	\omega_o = \sqrt{\frac{k}{M}}
\end{equation}
where $M$ is the mass of the object doing the motion (the piston) and $k$ is the spring constant related to the SHM and that we need to relate to the properties of the gas. 

\subsection{Adiabatic process and degrees of freedom}

An adiabatic process has no change in the total energy, heat does not flow into or out of the system. To minimise the heat flow in or out of the apparatus the cylinder and piston  are make of glass, a poor thermal conductor. The experiments themselves only last a few seconds at most. As Fourier's law states, any heat loss will be proportional to the temperature difference from the gas to the glass. Below we'll propose that only in the most extreme experimental conditions (very small volume) would there be a large enough temperature difference to drive substantial heat transfer. There is some suggestion of this non-adiabatic nature in the volume dependence of the lifetime at small volumes. 

If there is no heat transfer into or out of the gas system, then the work done by the piston on the gas (or vice versa) will be an adiabatic process. We therefore  start with the well know expression for an adiabatic process,
\begin{equation}
	pV^\gamma = \textrm{constant},
\end{equation}
where $\gamma$ is heat capacity ratio,
\begin{equation}\label{eq-adiabatic}
	\gamma = \frac{C_p}{C_V}
\end{equation}
between the constant pressure heat capacity $C_p$ and constant volume heat capacity $C_V$. Leading to the expression for $\gamma$ that depends on the identity of the gas particles through its number of degrees of freedom $f$,
\begin{equation}
	\gamma = 1 + \frac{2}{f}.
\end{equation}
We will discuss the significance of $f$ and its temperature and molecular dependence in the General Discussion section \ref{sec_GD}, but fundamentally it means that a measurement of macroscopic and thermodynamic property $\gamma$,  gives measurable insight into the structure and properties of the individual gas atoms or molecules. This is why this experiment is ideally suited to bridge the gap between scripted experiment and open-ended projects. It links easy to measure properties (the plunder bobbing up and down) to the microscopic, or quantum, properties of the individual atoms and molecules. 

To estimate the accuracy required to determine significant difference between the four gases measured, Table \ref{DOF} presents the number of degrees of translational and rotational freedom each gas has. Vibrational degrees of freedom are discussed in section \ref{sec_GD}. Here we assume air is predominately nitrogen and oxygen, both diatomic molecules. This simple analysis suggest $\gamma$ will have to be measured to at least an uncertainty of 10 \% and preferably closer to the 1 \% level. 

\begin{table}
\caption{\label{DOF}%
Translational and rotational degrees of freedom and associated $\gamma$ value for the four gases measured (does not include vibrations).}
\begin{indented}
	\lineup
	\item[]\begin{tabular}{| c | c | c | c |} 
		\hline 
		Gas & Translational $f$  & Rotational $f$ & $\gamma$ \\
		\hline
		He & 3 & 0 & 1.667\\
		Air & 3 & 2 & 1.4\\
		N$_2$ & 3 & 2 & 1.4\\
		CO$_2$ & 3 & 2 & 1.4\\
		\hline
	\end{tabular}
\end{indented}
\end{table}

\subsection{From adibatic process to Hooke's law}\label{sec-adiabatic}

At the initial conditions we have that,
\begin{equation}
	\textrm{constant} = p_0 V_0^\gamma,
\end{equation}
here $p_0 = p_\textrm{atm} + Mg/A$ is a combination of the atmospheric pressure $p_\textrm{atm}$ bearing down on the top of the piston and pressure due to the weight of the piston itself $M$, gravitational acceleration $g$ and the cross-sectional area of the piston $A$. Atmospheric pressure was measured at the beginning of each experimental session, see experimental section \ref{sec_ex}. Due to the weight of the piston being relatively lighter then the `weight' of the atmosphere, the pressure correction is negligible, $\sim 1$ kPa, compared with typical atmospheric pressure of $101$ kPa. 

To find the restoring force caused by a displacement we rearrange to give the pressure as a function of volume,
\begin{equation}
	p(V) = p_0\left( \frac{ V_0}{V} \right)^\gamma.
\end{equation}
A computer simulation could then compute this full expression, through for example the Verlet algorithm, at each piston position and so numerically find the time-dependence of the piston. However, here to allow a match to SHM we consider small displacements to so see how the pressure varies to 1st order - the linear approximation. A Taylor expansion of the pressure with a small volume change $\Delta V$ around the initial volume $V_0$ gives,
\begin{equation}
	p(V_0 + \Delta V)  \approx p(V_0) + \left. \frac{dP}{dV}\right|_{V=V_0} \Delta V \\
	+ \frac{1}{2}  \left. \frac{d^2P}{dV^2}\right|_{V=V_0} (\Delta V)^2 + \textrm{small terms}.
\end{equation}
We can see that $p(V_0) = p_0$, the next term
\begin{eqnarray}
	\frac{dP}{dV} &= -\gamma p_0 \frac{V_0^\gamma}{V^{\gamma + 1}} \\
	\therefore \left. \frac{dP}{dV}\right|_{V=V_0} &= - \gamma \frac{p_0}{V_0}.
\end{eqnarray}
Similarly 
\begin{equation}
	\left. \frac{d^2P}{dV^2}\right|_{V=V_0} = (\gamma^2 + \gamma) \frac{p_0}{V_0^2}.
\end{equation}
The total pressure difference across the piston, the equivalent of subtracting $p_0$, gives the pressure imbalance by a small change in volume as
\begin{equation}\label{pressure_taylor}
	p(\Delta V) \approx  -\gamma p_0 \left(\frac{\Delta V}{V_0}\right) + \frac{1}{2} (\gamma^2 + \gamma)p_0 \left(\frac{\Delta V}{V_0}\right)^2.
\end{equation}
Taking the 1st order, $\Delta V$, term and typical values for helium, at $100$ ml initial volume the maximum increase in pressure due to the oscillations will be $\approx 4$ kPs and at the smallest initial volume of $10$ ml a pressure increase of $\approx 40$ kPa. Well within the safety limits of the glass apparatus. 

Finally to mirror a Hooke's law spring, we convert to force
and setting $\Delta V = A \Delta z$ and $V_0 = z_0 A$ to give
\begin{equation}
	F(\Delta z) \approx  \underbrace{-\gamma p_0 A \left( \frac{\Delta z}{z_0} \right)}_{\textrm{Linear in } \Delta z}+ \underbrace{\frac{1}{2} (\gamma^2 + \gamma) p_0 A \left( \frac{\Delta z}{z_0}\right) ^2}_{\textrm{Quadratic in } \Delta z}.
\end{equation}
If we assume $\gamma \sim 1$ for all gas particles, we can see that for say a $1$ in $100$ deviation from a linear response to displacement from the equilibrium $z_0$, e.g., where the linear term is 100 times larger than the quadratic term, we need that $\Delta z / z_0 \sim 1/100$. The normal experimental procedure was to displace by about $5$ mm and release.  The apparatus is made so that $1$ mm of height corresponds to $1$ ml of volume. At $100$ ml volume the cylinder is $100$ mm high. Therefore any low volume measurements, well below $100$ ml the response could significantly deviate from a linear response. Very low initial volume measurements are, however, also tricky as  from an analysis of such data there is an increase in the damping (presumable coming close to the static friction regime) and so we are restricted to rather larger initial displacements that therefore may well drive our system away form a purely linear response system at lower overall volumes.

In the linear, or 1st order, regime we can identify that $F=-k \Delta z$ so that $k = \gamma p_0 A^2/V_0$ and with $\omega_0 = \sqrt{k/M}$ we get the relationship
\begin{equation} \label{wd-square}
	\omega_0^2 = 	 \left(\frac{p_0 A^2}{V_0  M} \right) \gamma .
\end{equation}
The pressure $p_0$ is fixed by the atmospheric pressure and the mass of the piston which is also fixed, as is the area $A$ of the piston. This means we can vary the initial volume $V_0$ and measure the response $\omega_0$ to find the constant of proportionality, the adiabatic ratio $\gamma$.

\begin{table}
	\caption{\label{tb-conditions}%
		Measured atmospheric pressure ($\pm 0.1$ kPa), and laboratory temperature ($\pm 0.5 \ ^\circ$C) when the experimental results  were recorded.}
	\begin{indented}
		\lineup
		\item[]\begin{tabular}{| c | c | c |} 
			\hline 
			Gas & Atm Pressure / kPa & Lab temp / $^\circ$C \\
			\hline
			He & 101.1 & 23.3 \\
			Air & 101.4 & 23.5 \\
			N$_2$ & 101.1 & 22.5 \\
			CO$_2$ & 101.7 & 23.5 \\
			\hline
		\end{tabular}
	\end{indented}
\end{table}

From the ideal-gas law and the adiabatic  relationship of equation \ref{eq-adiabatic} the gas temperature is given by
\begin{equation}
	T = T_0 \left( \frac{V_0}{V} \right)^{\gamma - 1}
\end{equation}
where $T_0$ is the initial temperature of the gas. For small volume changes around $V_0$ this can be Taylor expanded and so approximated as 
\begin{equation}
	\Delta T \approx T_0 \left( 1 - \gamma\right) \frac{\Delta V}{V_0}.
\end{equation}
The temperature change due to the oscillation will therefore be larger for gas with the larger $\gamma$ value (e.g. helium) and larger at the smallest volumes. For helium this will be $\Delta T \approx 9 \ ^\circ$C at $V_0 = 100$ ml and  $\Delta T \approx 90 \ ^\circ$C at $10$ ml. Whereas for carbon dioxide both are lower  $\Delta T \approx 4 \ ^\circ$C at $100$ ml and  $\Delta T \approx 41 \ ^\circ$C at $10$ ml. These results are broadly in line with, say, the same effect felt when pumping up tyres on a bike as the pump get hot to the touch. Therefore, any nonadiabaticity (driven by the larger temperature gradients from gas to external heat bath) will be more likely at the smaller volumes investigated. 

Although we did not do this, it would be interesting and instructive to directly measure the time-dependence of the gas temperature and pressure. From this students could try and use the displacement measurement to reconstruct of the temperature and pressure and compare that with the measured values. That is, not only using an equation to compute other parameters, but measure all the parameters to see if that equation holds. 

\subsection{diabtic SHM}

The same analysis for a purely diabatic process, that is one at constant temperature, gives the result $\omega_0^2 = p_0 A^2/V_0  M$, which means that if there is any heat flow from or to the apparatus it will naturally lead to a lowering of the measured value of $\gamma$ towards 1.

\section{Experimental}\label{sec_ex}

\subsection{Apparatus and procedures}

\begin{figure}
	\begin{center}
	\includegraphics[trim = 14cm 1cm 12.5cm 0cm ,clip = true, height=0.65\textheight]{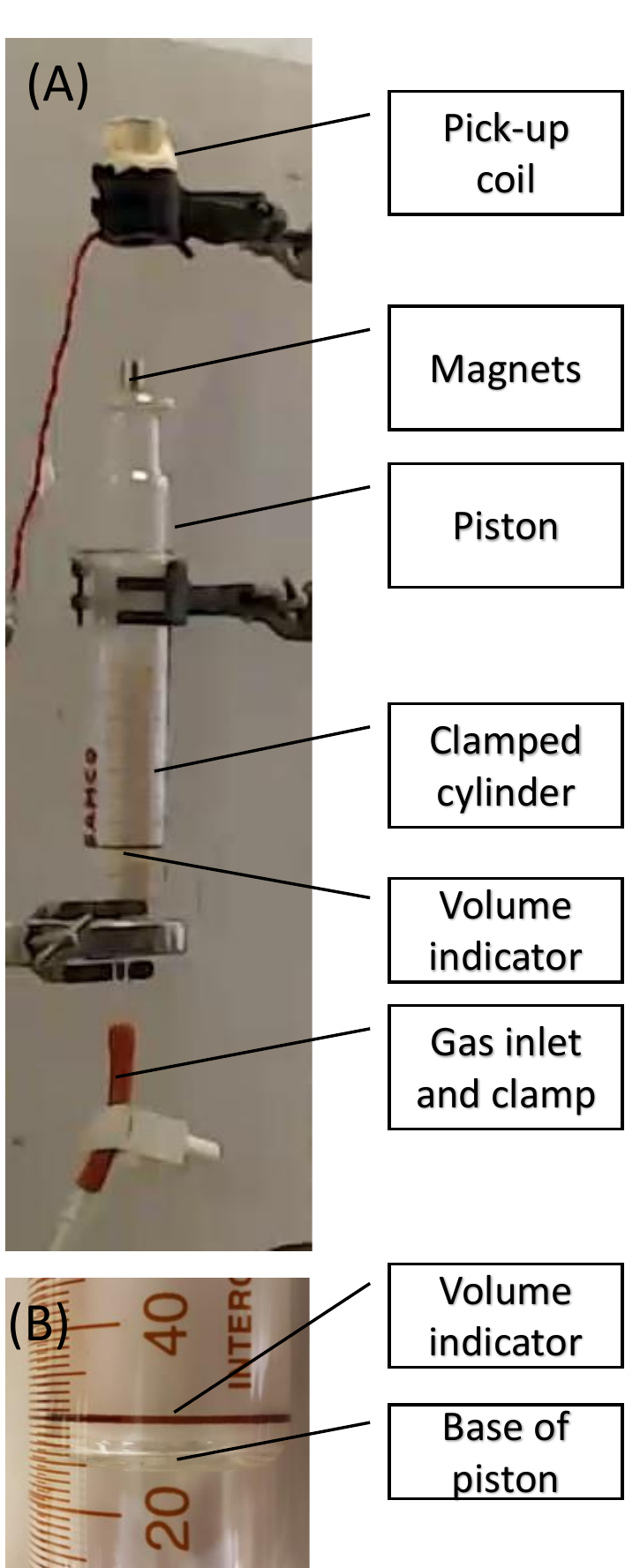}
	\caption{\label{apparatus} %
		(A) Photograph of the apparatus with labels. Gap between top of magnets and bottom of pick-up coils was 50 mm. (B) Detailed photographs of the   base of the piston and the gas tight seal.
	}
	\end{center}
	\end{figure}
The apparatus consisted of a glass cylinder and movable piston with a good gas-tight seal due to the exact fitting of the piston within the cylinder. This large glass-glass interface only introduced minimal frictional damping. To measure the motion of the piston three small rare-earth magnets were glued to the top of the piston. As the piston/magnets moved this induced an electromotive force (emf) in a pick-up coil $(50 \pm 5)$ mm above the magnets and was measured and captured by an oscilloscope. Figure \ref{apparatus} shows (A) a photograph of the whole apparatus and (B) a close up photograph of the end of the piston and the gas tight  seal.

To ensure a smooth operation and to align the magnet and coil a laser light was passed from above through the coil onto the top of the stack of magnets. This allowed adjustment of the clamps and orientation of the coil to align it the axis of the magnets as accurately as possible. Moreover, as the cylinder was filled with gas, so the starting point of the magnet would naturally rise. To keep a consistent average emf in the pick-up coil, the coil was at each gas-volume moved to ensure a 50 mm gap between top of the magnet stack and the base of the coil. This level of accuracy was only important if a full understanding of the magnitude of the emf signal was wanted but  isn't critical to the measurement of the adiabatic ratio $\gamma$.

A precise procedure for emptying, filling, purging and filling  the cylinder with pure gas was followed. We would estimate that perhaps there could be at most a 1 \% (ratio of volume of pipework open to air to volume of the maximum fill) contamination with air. This would obviously not effect the air measurements, nor impact on the nitrogen measurements, but could have some influence on the helium (lowering $\gamma$) and carbon dioxide (raising $\gamma$) measurement and there is some evidence of that.

To start the dSHM the piston was pushed down by hand by approximately 5 mm (reducing the volume by 5 ml) and released. The oscilloscope was set up to capture a single time-trace based on a trigger. Example time-traces are shown in figure \ref{time-trace}. Each gas was measured over 10 volumes from 10 ml to 100 ml, see for example figure \ref{volumes} and each volume had 5 repeat measurements. The volume measurements were taken by reading from the graduated scale of the cylinder the position of the black-seal ring. The uncertainty in all initial volume measurements was estimated to be $\pm 1$ ml. This is  the main factor limiting the accuracy of the measurement of the adiabatic ratio $\gamma$.

A careful measurement of the actual volume of the cylinder-piston system to the graduated values found that there was an additional 6 ml of volume to the marked volume. This volume was attributed to the gas inlet tube system and slight curvature in the base of the piston. Therefore for a reported volume of, say 50 ml, the true, or corrected, volume was 56 ml. For ease of reading this correction is only made to the final fitting function of figure \ref{gamma_and_f}.

The piston and magnets had a combined mass of  $M = (106.68 \pm 0.01)$ g, and the piston had a diameter of $(34.16 \pm 0.01)$ mm. The pick-up coil has a diameter of $(13 \pm 1)$ mm and $n = (1400 \pm 20)$ turns.  Gasses were of purity 99.8 \% or better. Local gravitational acceleration was calculated using the International Gravity Formula and using The University of Bath's latitude of 51.3811$^\circ$ and an elevation of the laboratory of 192 m above sea level giving $g = 9.81134$ ms$^{-2}$.

\subsection{Damping and gas leakage}

A weakness of this  approach and apparatus is the seal between cylinder and piston. It must be sufficiently gas-tight to prevent leakage (and hence change of volume) of the gas, yet allow the nearly free movement of the piston within the cylinder. The apparatus here is simply sealed by the precise fit of the piston within the cylinder.

At volumes larger than 50 ml we find damping lifetimes consistent at $\sim 1$ s (see figure \ref{fig_fitting_parameters}) which when compared with the typical dSHM frequency of $\omega_d \sim 130$ rad s$^{-1}$ results in only a minor correction to compute an undamped frequency, in this example, of $+1\times 10^{-3}$ rad  s$^{-1}$. Below an initial volume of 50 ml  there is a much larger damping effect. 
\begin{figure}
	\begin{center}
	\includegraphics[height=0.65\textheight]{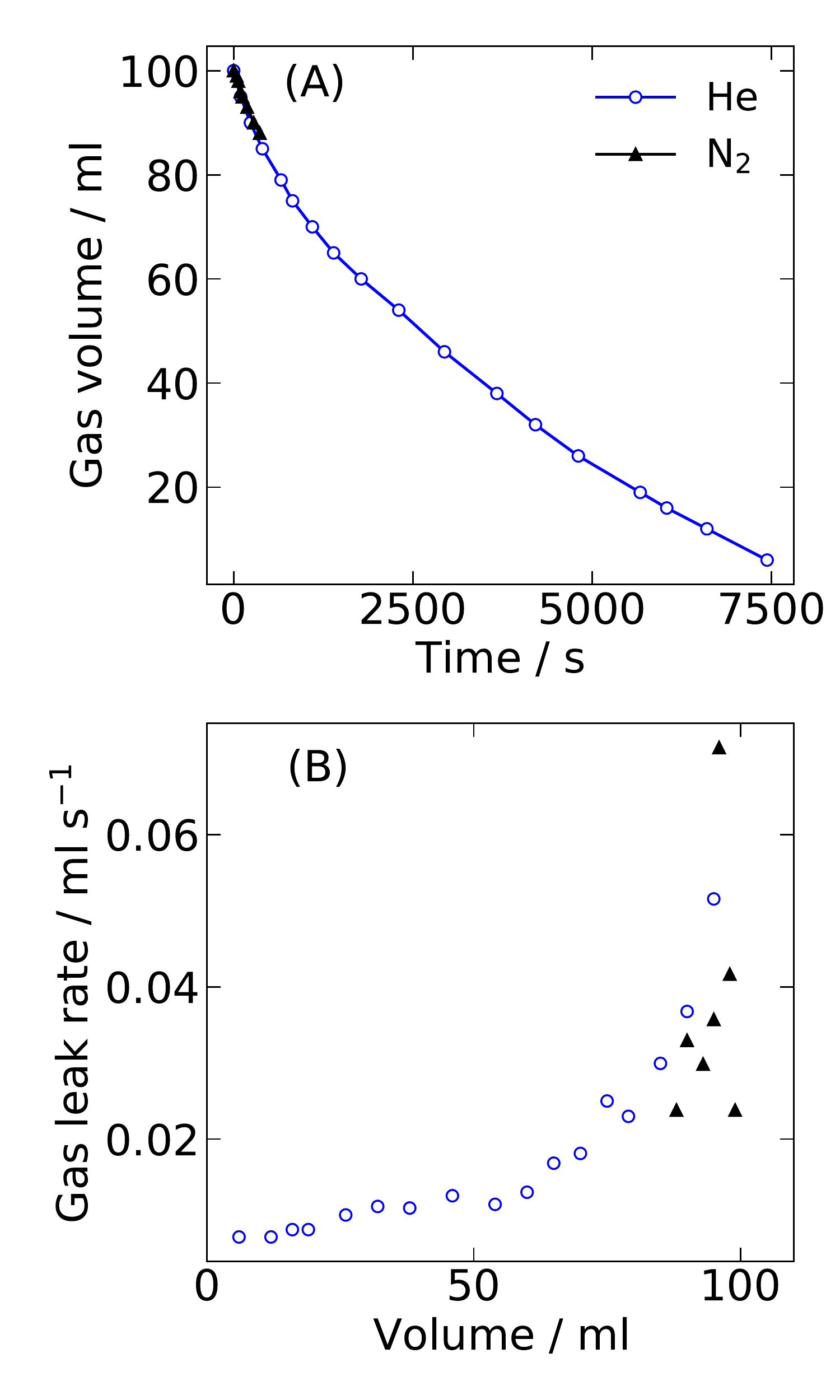}
	\caption{\label{leakrate} %
	Gas leakage rates from the cylinder-piston apparatus. (A) The time-dependence of volume of the system from an initial fill, of both Helium and of Nitrogen gas, to 100 ml as it naturally reduced due to the pressure imbalance due to the mass of the piston-magnet system. (B)  Pair-wise simple differential of (A) to show the gas leak rate.
	}
	\end{center}
\end{figure}

Figure \ref{leakrate} shows the volume of the apparatus filled with  helium and with nitrogen gas naturally changes over time. This is due to the mass of the piston itself producing a pressure difference of $\Delta p = Mg/A = 1.14$ kPa which although much lower than the mean atmospheric pressure of $101.3$ kPa is enough to drive gas form the piston through the seal. From this we determine the gas leakage rate as shown in figure \ref{leakrate}(B), this is just the simple neighbouring volume and time differences of the data of figure \ref{leakrate}(A). The leak rate  shows a similar low-volume and high-volume regime as the damping rate.  

This is unexpected. Given that the pressure difference between gas and atmosphere should be constant in this gas-leak experiment, and that the only change in the apparatus is the position of the seal we may expect a volume independent leakage rate. Therefore, we speculate that there is an increase in the friction between seal and cylinder below the 50 ml mark. This naturally leads to an increased, and observed, damping (decrease in the lifetime). Furthermore, there will be an increase in the friction opposing the piston as it slowly descends in the cylinder leading to a reduction in the pressure difference between the gas and atmosphere and so a reduction in the leak-rate.

What we can also see is during a typical oscillating experiment of say, 2 s, there will be at most a 0.1 ml loss of gas from the system therefore we can safely ignore such gas loss from the analysis of any one experimental time-trace measurement. In future experiment the volume before and after each dSHM run should be explicitly measured to quantify any gas leakage directly.

\subsection{Magnetic field}

There are of course many ways to measure the time-dependent movement of the piston. Here to ensure a non-invasive measurement, but based on introductory physics we make use of the electromotive force (emf) generated by a moving magnet. A set of small magnets glued to the top of the piston and a pick-up coil connected to an oscilloscope to measure the time-dependence of the induced electromotive force (emf) $\epsilon$. Through Faraday's law of induction 
\begin{equation}
	\epsilon = - \frac{\textrm{d}\Phi_B}{\textrm{d}t} \label{faraday}
\end{equation}
where $\Phi_B$ is the magnetic flux through some area, we can relate the measured signal to the piston dynamics. The area is the area $A_c$ of the pick-up coils multiplied by the number of turns $n$. If we take the on-axis description of the magnetic field from the dipole source of the magnets, we have $B(z)$ the magnetic field strength 
\begin{equation}
		B(z) = \frac{2\mu_0}{4\pi z^3}m_B,
\end{equation}
where $\mu_0$ is the vacuum permeability, $z$ is the distance away from the centre of the magnet and $m_B$ is the magnitude of the magnetic-dipole. Assuming the flux is the same for all the coils, in effect ignoring the height of the coil stack and the slight off-axis nature of the total area of the coils, we have 
\begin{equation}
	\Phi_B = A_c n\left( \frac{2\mu_0}{4\pi z^3} \right) m_B. \label{phib}
\end{equation} 
Here a simple 3D calculation using standard mathematical models the full dipole magnetic $B$-field could be done to convince a student that these approximations are valid.

\begin{figure}
	\begin{center}		
	\includegraphics[height=0.65\textwidth]{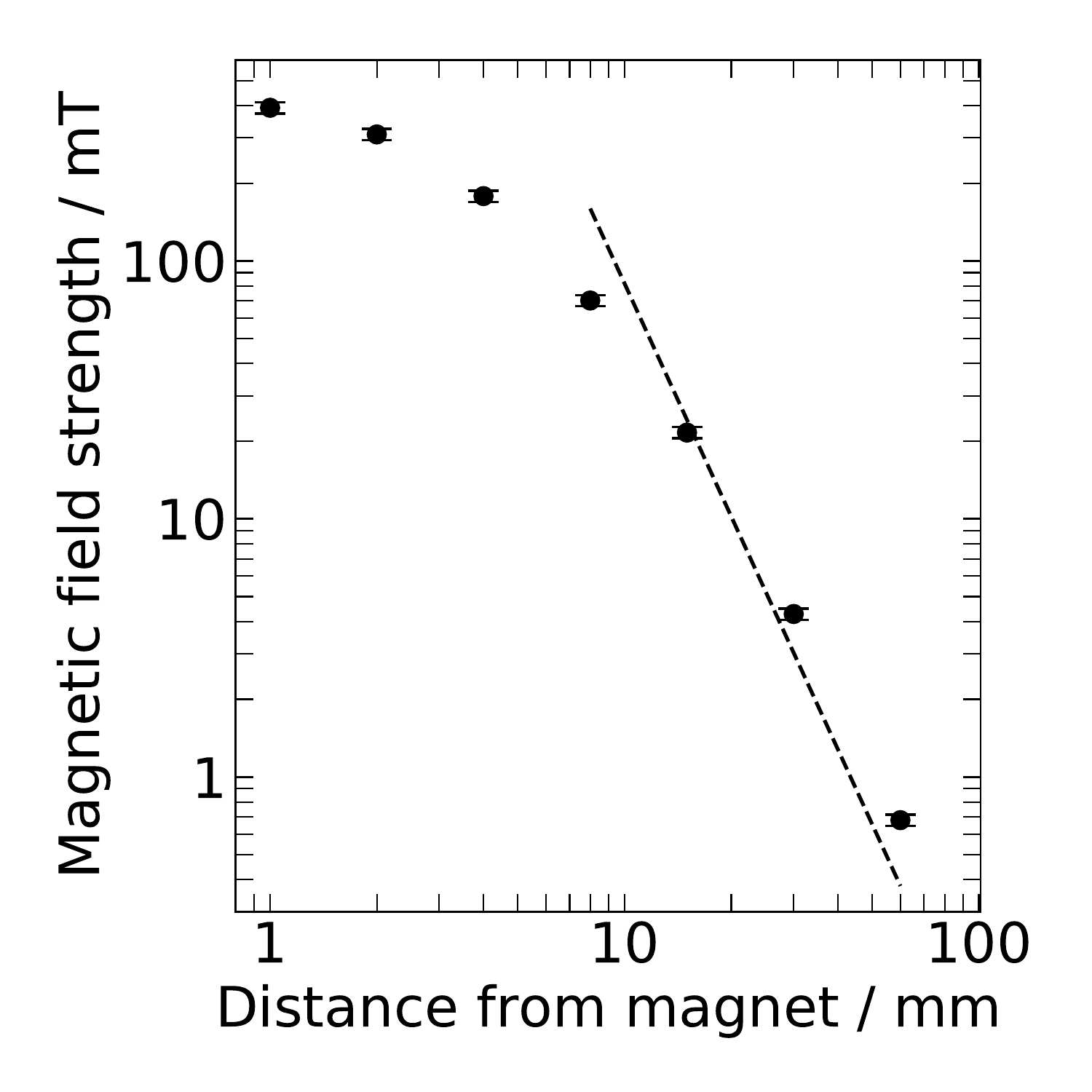}
	\caption{\label{magnet} %
		Magnetic field strength as a function  of distance - note log-log scale. See main text for details of fitting function that was fitted across data as shown.
	}
	\end{center}
\end{figure}

To find a value for $m_B$ the magnetic field strength was measured using a Hall probe at several distances from the magnet as shown in figure \ref{magnet}. 
On this log-log graph we would expect a linear dependence, which is evident at distances greater than 10 mm. As the experimental gas data used an initial $z_m = 60$ mm separation (approximate midpoint of coil height) with a $z_0=5$ mm initial SHM amplitude, we only consider this region. (At lower distances the magnetic field measurement is probably sensitive to slight misalignment of the Hall probe both off-axis and off angle.) To ensure an equal weighting of the data within the usual least-square-fitting of SciPi's `curvefit' module, we first took the logarithm of data to fit the expression $ y = \log_{10}(c) - 3 x$ and find $m_B = (0.41 \pm 0.13)$ Am$^2$.

The time-dependence of the measured signal emf $\epsilon$ is therefore introduced by the time-dependence of the position of the magnet as it oscillates relative to the pick-up coil with dSHM, this is in essence combining equations \ref{damped-SHM}, \ref{faraday} and   \ref{phib}. To solve this fully, the $B \propto 1/z^3$ dependence would need to be taken into account. (If that dependence is taken into account a more sawtooth like emf waveform is computed from simple simulation and is consistent with some of our early measurements where we used too large and initial amplitude and at too close a position.) Here we make the further approximation that the displacement caused by the dSHM is small in comparison with the overall distance between magnet and coil. Taking a Taylor expansion to first order of equation \ref{phib} about the initial $z_m=50$ mm position gives
\begin{equation}\label{phib-taylor}
		\Phi_B(z_m + \Delta z) = \left( \frac{m_B\mu_0A_c n}{2\pi z_m^3} \right)  - \Delta z \left( \frac{m_B 3\mu_0A_c n}{2\pi z_m^4} \right).
\end{equation}
Now combining  equations \ref{damped-SHM}, \ref{faraday} and \ref{phib-taylor} for the dSHM, Faraday's law and the magnetic flux $z$-dependence gives
\begin{equation}
\epsilon(t) =  \left( \frac{m_B 3\mu_0A_c n}{2\pi z_m^4} \right)  \frac{\textrm{d}}{\textrm{d}t}z(t)
\end{equation}
and so
\begin{equation} \label{emft}
	\epsilon(t)	=\beta \left[ \omega_d \cos \left(\omega_d t + \phi \right) - \frac{1}{\tau} \sin \left( \omega_d t + \phi \right) \right] \exp \left(-t/\tau\right)
\end{equation}
where we combine all the constants into one value $\beta = m_B 3\mu_0A_c n z_0 / 2\pi z_m^4$. The maximum emf will be just after at $t=0$ where $\omega_dt + \phi = \pi/2$. That would give a predicted maximum emf of  $\sim 4$ mV which matches well the actually measured values (see fig. \ref{time-trace}) of  $\sim 6$ mV.  Note that to measure the gas properties, as the relationship of equation \ref{wd-square} shows, the absolute magnitude of the emf is not  required only the angular frequency. 

As the signal looked like dSHM and not the slightly more nuanced expression of eq \ref{emft}, most students struggled with the linkage of the dSHM through the magnet/coil system  using Ampere's law.   
		
\section{Results and Specific Discussion}

\begin{figure}
	\begin{center}
	\includegraphics[width=\textwidth]{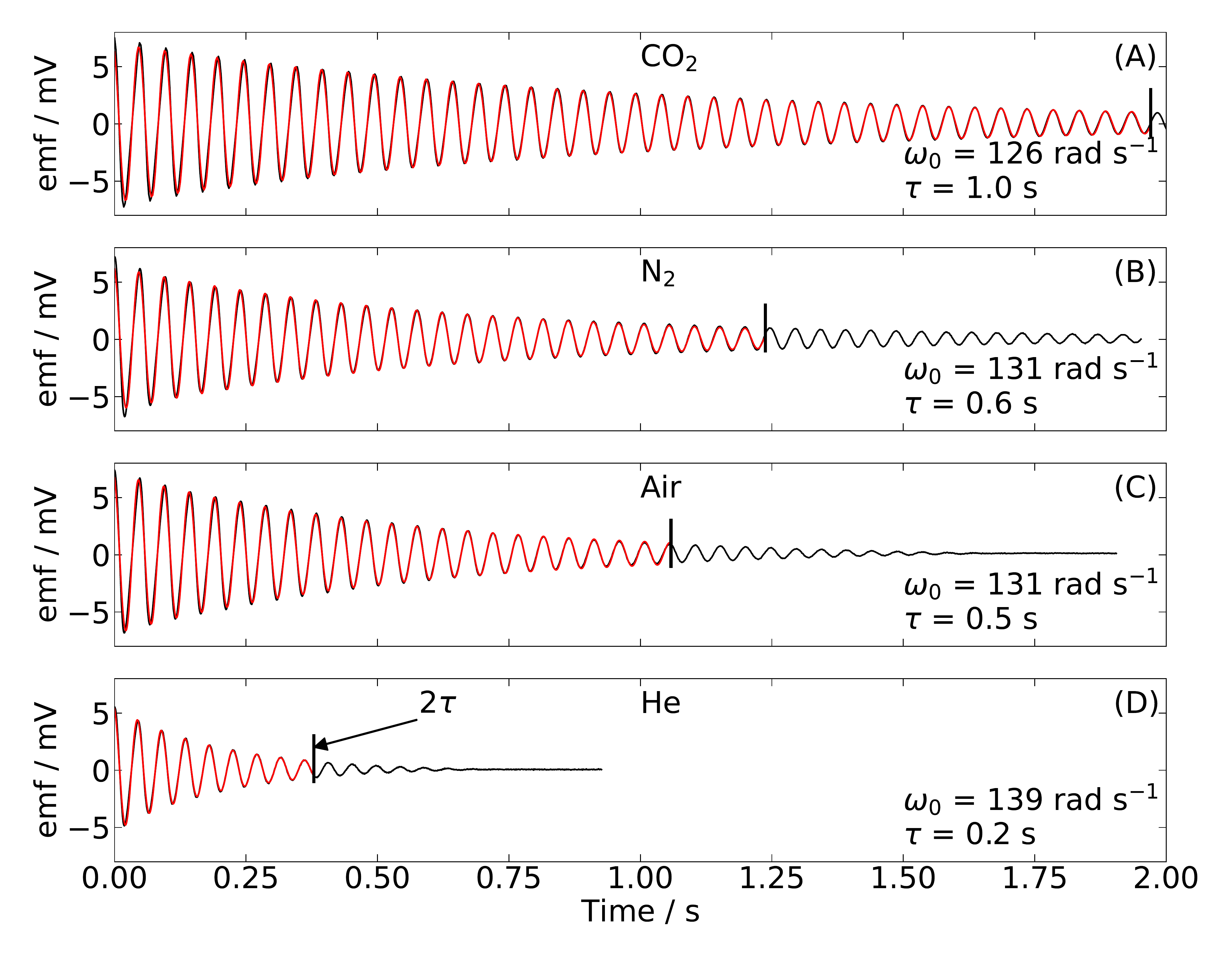}
	\caption{\label{time-trace} %
		Typical raw oscilloscope voltage time-traces of the emf induced in the pick-up coil generated by the dSHM of the piston containing:(A)  Carbon-dioxide , (B) Nitrogen, (C) Air and (D) Helium. The (nearly perfectly) overlaid red line is a dSHM model as described in the main text fitted from $t = 0$ to $t = 2\tau$. The short black vertical line marks the $t = 2\tau$ position. The extract values of the undamped angular frequency and decay lifetime as given for each time-trace. Each experiment was at an initial volume 60 ml and other parameters as in the Experimental section. Uncertainties in fitted value of $\omega_0 \sim \pm 3$ mrad s$^{-1}$, uncertainties in $\tau \sim \pm 7$ ms.  
	}
	\end{center}
\end{figure}

Figure \ref{time-trace} shows an example emf time-trace, at an initial volume 60 ml, for all four gasses, (A) carbon dioxide, (B) nitrogen, (C) air and (D) helium. Each has approximately the same initial emf value as they were all started by a 5 mm compression and so have the same initial conditions.   

All time-traces display the same main features: that there is an oscillation of approximately the same frequency, and an overall decay of that oscillation. To extract precise values for these parameters the SciPi package curvefit was used with  equation \ref{emft} with fitting parameters $\beta, \omega_d, \phi$ and $\tau$. Since each time-trace experiments was started by hand and therefore there will be a natural range of initial $z_0$ initial displacement, and since the absolute magnitude of the emf is not required to find the adiabatic gas parameters, all the amplitude terms were simply combined into the one fitting parameter $\beta$. The (near perfectly) overlaid red curves in figure \ref{time-trace} are best fits of this model to the various data sets. The fitted  values of $\omega_d$ and $\tau$ are displayed on the relevant plots.

To allow automated and unbiased fitting, the following  procedure coded into a python script. The maximum emf value was found and this point was set at $t=0$, all data points before that were removed from the fitting data-set. The lowest (most negative) data point was found, and by comparing that time with the time of the maximum emf value an approximate value of $\omega_d$ was determined. To determine an approximate value for $\tau$ a simple single exponential decay was fitted to the absolute value of the emf signal as a function of time. Finally using these parameters an initial value of $\beta$ was determined to ensure the $t=0$ value of equation \ref{emft} returned the correct $t=0$ value of the measured signal. Using these initial inputs for the curve-fitting routine equation \ref{emft} was fitted to the data-set. 

It was clear (not shown here) that at very low amplitude, below 1 mV, there was a decrease in the damping life-time. At these very low amplitudes it maybe that there is a increase of the damping friction as the system begins to feel the effect of static friction. Therefore, finally the same model was fitted again, but using the best-fit parameters found from that first fit and only fitting from $t=0$ to $t=2\tau$. This can be seen in the extent of the red-line fit and vertical bar for each of the data-sets shown in figure \ref{time-trace}.

\begin{figure}
	\begin{center}
	\includegraphics[height=0.85\textheight]{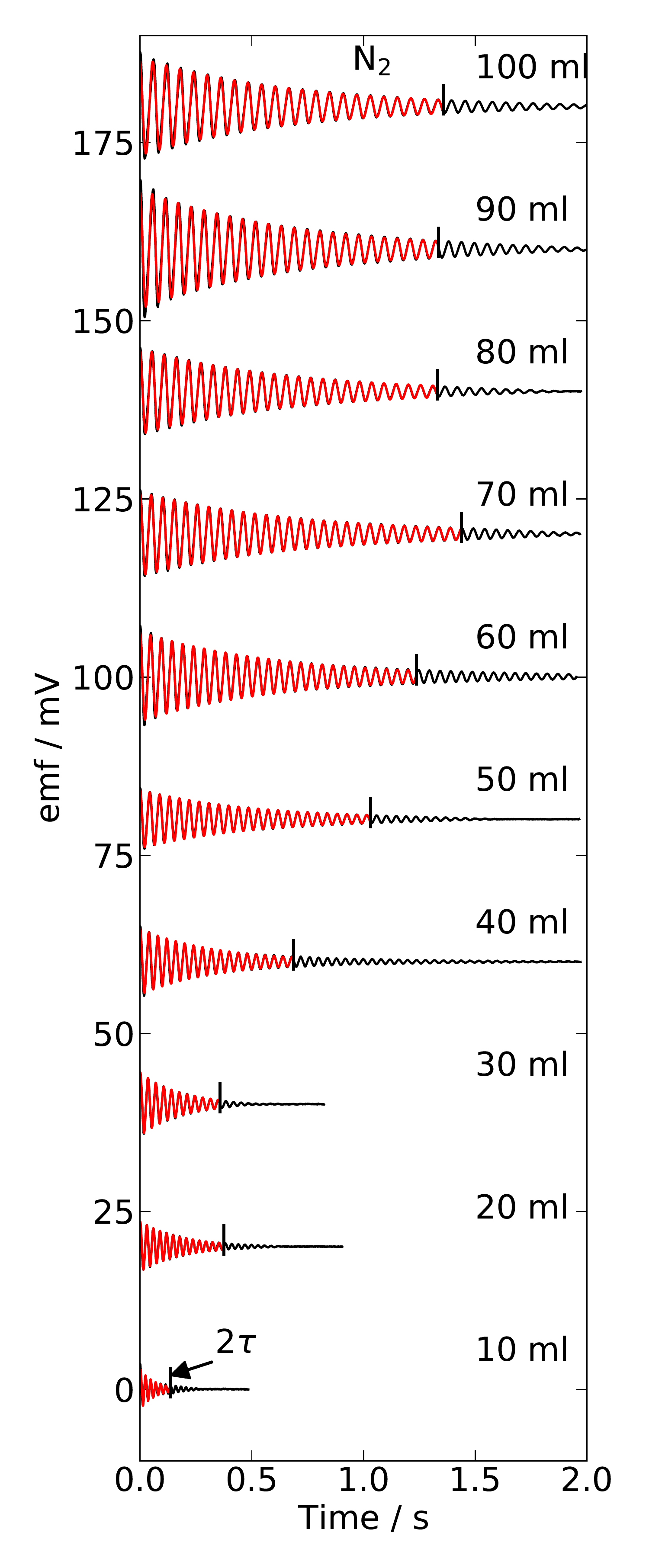}
	\caption{\label{volumes} %
		Series of typical emf time-traces for nitrogen gas over 10 cylinder volumes from 100 ml to 10 ml as indicated. To aid clarity each has been shifted vertically by 20 mV. The overlaid red line is, again, a line of best fit to a dSHM model that was fitted over a time scale of $2\tau$ for each individual time-trace. The short black vertical line marks the $t = 2\tau$ positions.
	}
	\end{center}
\end{figure}

Figure \ref{volumes} shows emf time-traces for nitrogen gas for all 10 measured initial volumes ranging from 100 ml to 10 ml in 10 ml steps. The (near perfectly) overlaid red line of the fitting function shows that the model developed culminating in the expression of eq. \ref{emft} accurately describes the experiment across all gasses and experimental parameters. What is evident (for all gasses) is a near constant decay lifetime from 100 ml to 60 ml followed by a dramatic reduction below that volume. There is also an obvious increase in the rate of oscillation as the initial volume decreases exactly in line with th expectation of the dSHM and adiabatic gas properties derived in section \ref{sec-adiabatic}. As the initial volume shrinks, so the adiabatic compression results in a stiffer response (spring constant) and hence higher oscillating frequency.

\begin{figure}
	\centering
	\includegraphics[height=0.85\textheight]{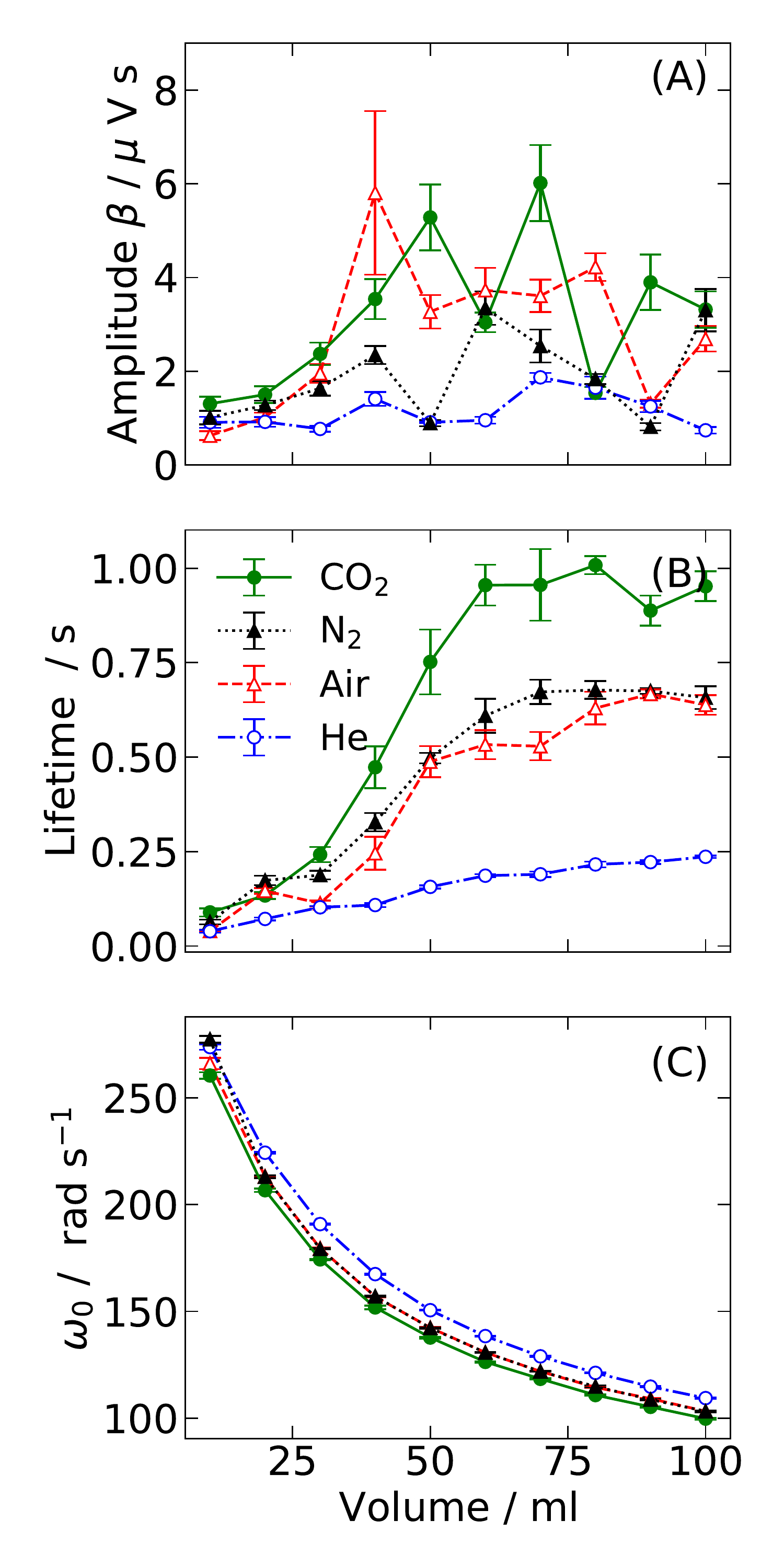}
	\caption{\label{fig_fitting_parameters} %
	Gas and volume dependence of the parameters (A) emf amplitude $\beta$, (B) decay lifetime $\tau$ and (C) undamped angular frequency $\omega_0$. Each point and associated error bar is the average over 6 time-traces with parameters taken from the best fit of equation \ref{emft}. Gases are: carbon dioxide, green solid line; nitrogen black dotted line; air, red dashed line; and helium, blue dash-dot line.   
	}
\end{figure}

Figure \ref{fig_fitting_parameters} presents the three fitting parameters, amplitude of emf, decay lifetime and undamped angular frequency, for all four gasses across all volumes. Each data point and associated error bar is the result of averaging across six time-traces for each set of experimental parameters.  
The emf amplitude $\beta$, shown in panel (A), which is proportional to the initial displacement $z_0$, is remarkably constant given the human instigation of the oscillations. At very small volumes it becomes quite difficult to depress the piston by hand and this may lead to an inadvertent reduction of the amplitude at  low ($V < 25$ ml) volumes. 

What is striking is the volume dependence of the decay lifetime as shown in Fig. \ref{fig_fitting_parameters}(B). It is reasonably constant and high for all gasses for initial volumes in the range $100$ ml to $50$ ml. Below $50$ ml, however, there is a stark reduction of the lifetime. This is the equivalent of an increase in the damping friction. This friction to volume dependence, is similar to that proposed to explain the earlier gas leakage rate measurements. It is therefore likely that there is a significant difference in the friction between cylinder and piston within the apparatus for the piston below or above the 50 ml volume mark. 

The damping lifetime seen in figure \ref{fig_fitting_parameters} shows a clear trend with carbon dioxide having the longest lifetime (least damping), air and nitrogen almost identical and helium the shortest lifetime (most damping). For a dSHM system the quality $Q$-factor can be related to the energy loss per cycle $\Delta E$ and the damping lifetime by
\begin{equation}
	Q = 2\pi \frac{E}{\Delta E} = \frac{\omega_0 \tau}{2}
\end{equation}
giving,
\begin{equation}
	\tau = \frac{4 \pi E}{\omega_0 \Delta E}
\end{equation}
where $E$ is the total energy of the SHM system. It is therefore possible to analyse the time-traces by Fourier analysis \cite{Caccamo_2019} to extract both the central frequency $\omega_0$ and the full width half max of the frequency domain response to extract $\tau$. Care should be taken with such analysis as the emf measurement is not simply dSHM, but its differential.

The usual assumption is that the damping is due to mechanical frictional  that will be constant across gasses. Another analysis \cite{Bringuier_2015} suggests that the energy dissipation is dominated by heat transfer (nonadiabaticity) of the apparatus. This is somewhat in line with the thermal conductivity (at $300$ K) of helium $156$ mW m$^{-1}$ K$^{-1}$ is much greater than that of nitrogen $26$ mW m$^{-1}$ K$^{-1}$, air $26.4$ mW m$^{-1}$ K$^{-1}$ and carbon dioxide $17$ mW m$^{-1}$ K$^{-1}$ \cite{CRCthermal}. Therefore any nonadiabatic effects would be more pronounced with Helium. It would be interesting to try heavier Nobel gasses, still with $f=3$ but with thermal conductivities closer to the other gasses studied, e.g., Argon $18$ mW m$^{-1}$ K$^{-1}$. This may explain why helium's measured $\gamma$ is lower than expected as any nonadiabatic processes will skew the measured value of $\gamma$ towards 1. Although that can also be the case for any contamination by air etc. 

The gas and volume dependence of  undamped angular frequency is presented in figure \ref{fig_fitting_parameters}(C). The rather long lifetimes, relative to the typical period of oscillation, for volumes at or above 50 ml means there is little difference, typically less then $+1$ rad s$^{-1}$, between damped and the corrected undamped angular frequency, see eq. \ref{damped_correction}. The frequency of oscillation was higher for helium at all volumes, nitrogen and air were closely matched, while carbon dioxide always had the lowest frequency. To restate for convenience eq. \ref{wd-square}  
\begin{equation} \label{wd-square2}
	\omega_0^2 = 	 \left(\frac{p_0 A^2}{V_0  M} \right) \gamma ,
\end{equation}
where the dependence of the angular frequency on the identity of the gas must be due to difference in the adiabatic ratio $\gamma$ and so by extension the number of degrees of freedom of the individual gases. As expected, see table \ref{DOF}, helium should have the highest $\gamma$ and so highest $\omega_0$, nitrogen and air should be near equal and lower than helium. 

In simple terms, the  work done by the piston moving a certain distance is, for helium, all converted into kinetic energy of the atoms as there are no other degrees of freedom. This leads to the largest pressure increase which is the restoring force of the SHM like spring system of this  R\"uchardt experiment. For nitrogen and air, that mechanical work is stored as both kinetic energy and rotational energy. Assuming an ideal gas like gas-gas (lack of) interactions, pressure is dependent on the kinetic energy, not rotational energy, therefore for those gasses the same piston displacement will result in a smaller pressure response and hence appear as a softer-spring and associated reduced frequency of oscillation. This kinetic and rotational only analysis suggests that carbon dioxide should be identical to nitrogen and air  - this is obviously and  consistently not the case. 

\begin{figure}
	\centering
	\includegraphics[height=0.83\textheight]{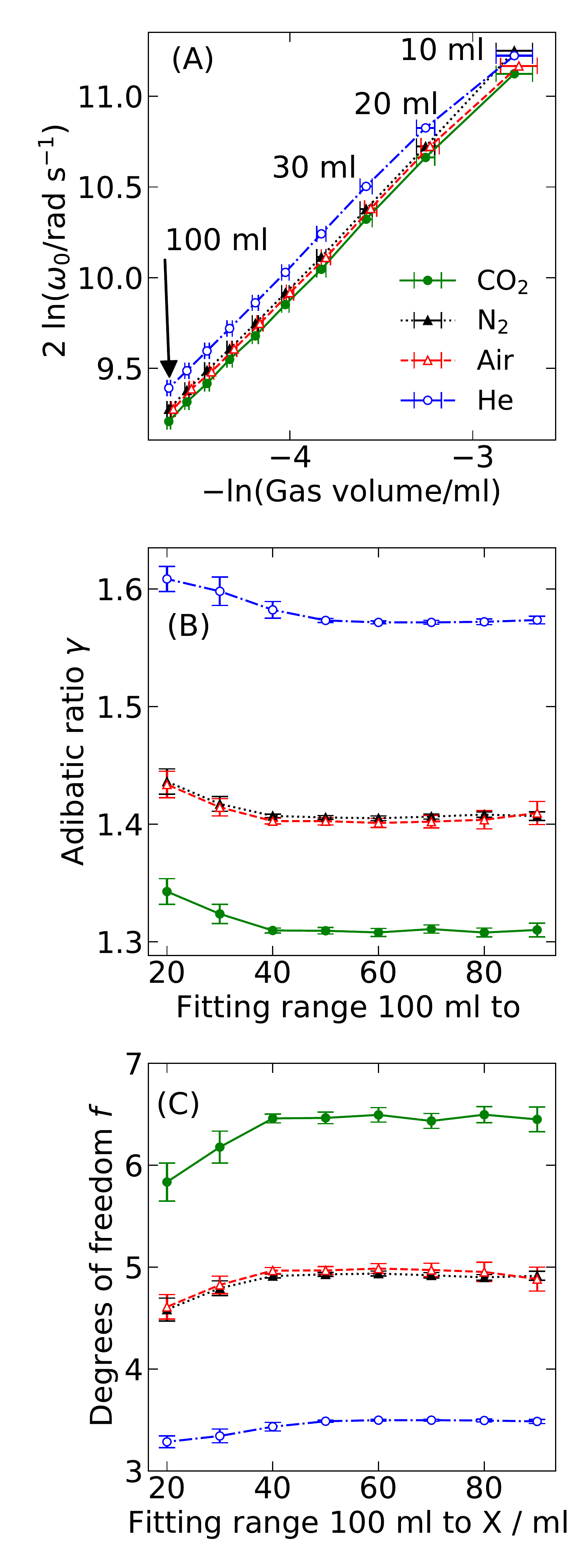}
	\caption{\label{gamma_and_f} %
	(A) The same data as figure \ref{fig_fitting_parameters}(C) but here with a log-log scale to demonstrate the linear and nonlinear dependence of the frequency of oscillation and the true volume (an additional 6 ml). (B) The extracted fitting parameter $\gamma$ found from fitting to figure \ref{fig_fitting_parameters} across ranges of volumes: smallest range from 100 ml to 90 ml, largest range from 100ml to 20 ml. (C) The number of degrees of freedom $f$ determined from (B). Note for ease of viewing, the data for Air has been slightly shifted in volume so as to not completely overlap with the nitrogen data.
	}
\end{figure}

To quantitatively determine accurate values for $\gamma$ for each gas figure \ref{gamma_and_f}(A) shows the same angular frequency data as figure \ref{fig_fitting_parameters}(C), but now with the true initial volume that includes the extra 6 ml offset due to the gas tubing etc. also, note the log-log scale.
It is clear that as the initial volume decreases, there is a deviation from the linear relationship found at larger volumes. This is plain in the fitted value of $\gamma$ shows in figure \ref{gamma_and_f}(B) where the fitting was performed over a range of experimental results from 100 ml to 90 ml, to 100 ml to 20 ml (the 10 ml data was dramatically more nonliner than the 20 ml data and so omitted or all analysis). Figure \ref{gamma_and_f}(C) shows the number of degrees of freedom $f$ associated with each $\gamma$ value. For the range of fitting up to and including the 50 ml value, there is a near invariant value for $\gamma$ (and hence $f$). This range matches the range of volumes found earlier that do not suffer large damping effects, or possible large nonadiabatic effects. We therefore take our best values as those found from fitting from 100 ml to 50 ml data, table \ref{tab-results} shows these values and the values that would have been attained with and without the $6$ ml correction to the volume. 
\begin{table}
		\caption{\label{tab-results}%
			Adibatic ratio $\gamma$ and uncertainty found from best fit over $100$ to $50$ ml data, for both the corrected volume that incudes 6 ml offset, and the uncorrected volume.}
		\begin{indented}
			\lineup
			\item[]\begin{tabular}{ c | c | c || c | c } 
			\hline 
			 Gas & \multicolumn{2}{c}{Corrected volume} & \multicolumn{2}{c}{Uncorrected volume} \\
			 & $\gamma$ & $f$  &  $\gamma$ & $f$ \\
			\hline
			CO$_2$  &  $1.3093(27)$ & $6.465(57)$ &  $1.196(10)$ &  $10.16(51)$  \\
			N$_2$  & $1.4057(16)$ &  $4.929(19)$ & $1.284(12)$ &  $7.02(29)$ \\
			Air  & $1.4025(32)$ & $4.968(39)$ &  $1.281(12)$ &   $7.09(29)$\\
			He  &$1.5733(17)$  &  $3.488(10)$&  $1.438(14)$ &  $4.56(14)$ \\
			\hline
		\end{tabular}
	\end{indented}
\end{table}

What is striking is the sensitivity to the volume that the number of degrees of freedom $f$ has. As this is a systematic uncertainty, or offset, it is not accounted for by the normal uncertainty analysis of the various linear least square fitting routines.  To determine an  error for this volume uncertainty, the same corrected-volume data was fitted again but with an additional correction volume of $+1$ ml or of $-1$ ml. The half difference of the final $\gamma$ and $f$ values from these fitting was  used to determine an uncertainty in the  original fitted values of $\gamma$ and $f$, see table \ref{tab-resultsError}.

\begin{table}
		\caption{\label{tab-resultsError}%
			Best adibatic ratio $\gamma$ found from fits over $100$ to $50$ ml data with estimation for a 1 ml systematic uncertainty in the corrected volume.}
		\begin{indented}
			\lineup
			\item[]\begin{tabular}{ c | c | c } 
			\hline 
			Gas & $\gamma$ & $f$   \\
			\hline
			CO$_2$  &  $1.309(19)$ & $6.46 \pm 0.39$   \\
			N$_2$  & $1.406(20)$ &  $4.92 \pm 0.24$  \\
			Air  & $1.403(20)$ & $4.96 \pm 0.25$ \\
			He  &$1.573(22)$  &  $3.48 \pm 0.14$ \\
			\hline
		\end{tabular}
\end{indented}
\end{table}

\section{General Discussion}\label{sec_GD}

Heat capacities, from the standard ref  \cite{CRCheatcap}, at 300 K of the  gasses are presented in table \ref{tab-crc}.
\begin{table}
		\caption{\label{tab-crc}%
			Literature values  for the adiabatic ratio for gasses at 300  K\cite{CRCheatcap}.}
		\begin{indented}
			\lineup
			\item[]\begin{tabular}{ c | c | c } 
			\hline 
			Gas & $\gamma$ & $f$   \\
			\hline
			CO$_2$  &  $1.293$ & $6.83$   \\
			N$_2$  & $1.400$ &  $5.00$  \\
			Air  & $1.402$ & $4.98$ \\
			He  & $1.667$  &  $3.00$ \\
			Ar  & $1.667$  &  $3.00$ \\
			 C$_3$H$_8$ & $1.127$  &  $15.78$\\
			\hline
		\end{tabular}
	\end{indented}
\end{table}
For the four studied gasses the measured experimental values from this work agree remarkably well with the known literature values. Nitrogen and Air are well within the experimental uncertainty. The measured value for carbon dioxide is lower but still within the uncertainty. Only Helium appears outside its expected range. As the heat capacity of helium has no temperature dependence (except near $4$ K) we can assign this difference to a combination of air contamination, and perhaps a large nonadiabatic effect.  

Heat capacity is dependent on the number of degrees of freedom a gas particle has. All gasses have three translational degrees of freedom, each mode counting $R/2$ to constant volume molar heat capacity $C_{V,m}$. Some have rotational degrees of freedom up to a maximum of 3, each mode adding $R/2$ to  $C_{V,m}$.  Some have vibrational degrees of freedom with the each mode adding $R$ to  $C_{V,m}$. The constant volume molar heat capacity of a gas is therefore expressed as the sum,
\begin{equation}
	C_{V,m} = \frac{1}{2} f R = \frac{1}{2}\left(3 + n_R + 2n_v \right) R,
\end{equation}
hence,
\begin{equation}
	 f =3 + n_R + 2n_v,
\end{equation}
where $R$ is the Universal gas constant, $n_R$ is the number of rotational degrees of freedom and $n_v$ is the number of vibrational degrees of freedom. 

If the moment of inertia of a rotation is dependent on the size of the atomic nucleus ($\sim 5 \times 10^{-4}$ \AA), rather than the length of a chemical bond ($\sim 2$ \AA) then quantum mechanics dictates that that axis of rotation will be blocked at room temperature. That is, if the energy level spacing of the rotational energy states is much larger than the equipartition of energy $k_BT/2$ that mode of rotation will be inaccessible to the molecule and so cannot effectively store energy - it is blocked - and does not contribute to the heat capacity.  Helium has all $3$-axes of rotation blocked, the linear molecules nitrogen and carbon dioxide will have one axis blocked and $2$ free to rotate.  As air is predominately nitrogen and oxygen, both linear diatomic molecules, on average Air will have only two modes of rotation. This analysis explains the measured values of $f$ for helium, nitrogen and air, but carbon dioxide, a triatomic linear molecule therefore also  with $n_R=2$, is higher than $5$ suggesting it has $n_v = 0.9$.

Like rotations, molecular vibrational only contribute to the heat capacity when there is enough thermal energy to excite them. Generally at room temperature this energy is insufficient to allow the molecule to explore its full vibrational spectrum. The normal equipartition of energy assumption of $k_BT/2$ will therefore not be valid, instead we must compute the partition function for each vibration in order to compute a vibration specific $n_v$. 

Carbon dioxide has four normal modes of vibrations, two energetically identical bending modes, a symmetric stretch and an asymmetric stretch, see Table \ref{tab-vibraitons}. For small vibrational amplitudes we can consider each an example of the quantum harmonic oscillator (another example of the Taylor expansion) with a characteristic vibrational energy ladder given by
\begin{equation}
	E_n = \left( n + \frac{1}{2} \right)  \hbar \omega,
\end{equation}
where $n$ is the quantum number of the vibration, $\hbar=$ Plank's constant$/2\pi$ and $\omega$ is an angular frequency characteristic of the vibration. For a stiff vibration this will be high, resulting in more widely spaced vibrational states in energy, and for a softer vibration the states will be closer together. Therefore for a fixed temperature, here 300 K, it is the relative scale of that vibrational ladder to the equipartition of energy that will dictate if that vibration adds, and by how much, to the number of degrees of freedom and hence the adiabatic ratio. The standard textbook of chemical information, the CRC Handbook of Chemistry and Physics, has tables of these parameters, but as it is written for a chemistry audience, physics students may struggle in converting from, usually a parameter in cm$^{-1}$ to the more typical eV energy unit. From tables \cite{CRCvib1} and \cite{CRCvib2} of the CRC Handbook we can find the information required.

 The diatomic molecule nitrogen has only  one vibrational mode, namely the symmetric stretch. 
\begin{table}
		\caption{\label{tab-vibraitons}%
			Details of the vibrational modes of carbon dioxide and nitrogen gasses. 
			Schematic digrams of vibrations, the vibrational excitation energy ($\Delta E$), their partitions $q^V$ function at 300 K and the corresponding number of degrees of freedom $n_v$ \cite{CRCvib1,CRCvib2}.}
		\begin{indented}
			\lineup
			\item[]\begin{tabular}{ c | c | c | c | c | c} 
			\hline 
			Gas & Vibration & Diagram & $\Delta E$ / eV  & $q^V$ & $n_v$\\
			\hline
			CO$_2$  &  Bend &\includegraphics[width = 3cm]{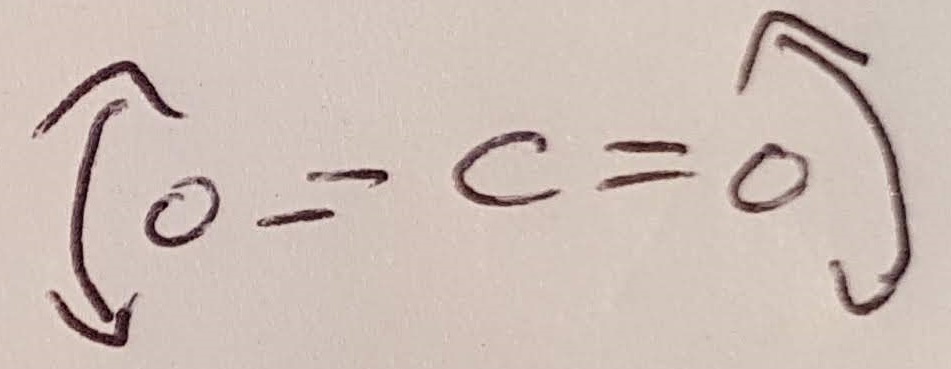} & 0.0827 & 1.043 & 0.454\\
			\hline
				CO$_2$	& Sym  &\includegraphics[width = 3cm]{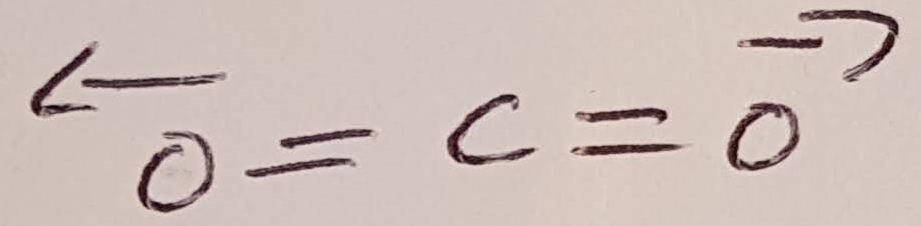} & 0.165& 1.002 & 0.069\\
					\hline
			CO$_2$		& Asym &\includegraphics[width = 3cm]{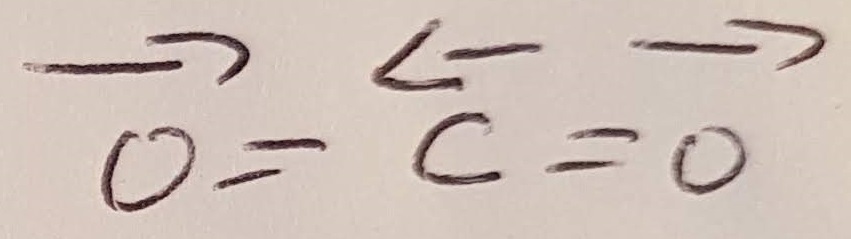} & 0.291 & 1.000 & 0.002\\
					\hline
					\hline
					
			N$_2$ & Sym &\includegraphics[width = 3cm]{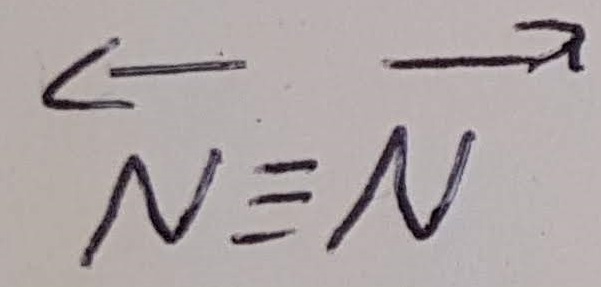} & 0.292 & 1.000 & 0.002\\
			\hline
		\end{tabular}
	\end{indented}
\end{table}
For each mode the partition function $q^V$ was computed via
\begin{equation}
	q^V = \frac{1}{1 - \exp{\left(-\Delta E / k_BT \right)}}
\end{equation}
at 300 K and the corresponding number of modes,
\begin{equation}
	n_v = \left( \frac{\Delta E}{k_B T} \right)^2 \frac{\exp{\left(-\frac{\Delta E}{k_BT}\right)}}{\left[1-\exp{\left(-\frac{\Delta E}{k_BT}\right)}\right]^2}.
\end{equation}
\begin{figure}
	\centering
	\includegraphics[width=0.65\columnwidth]{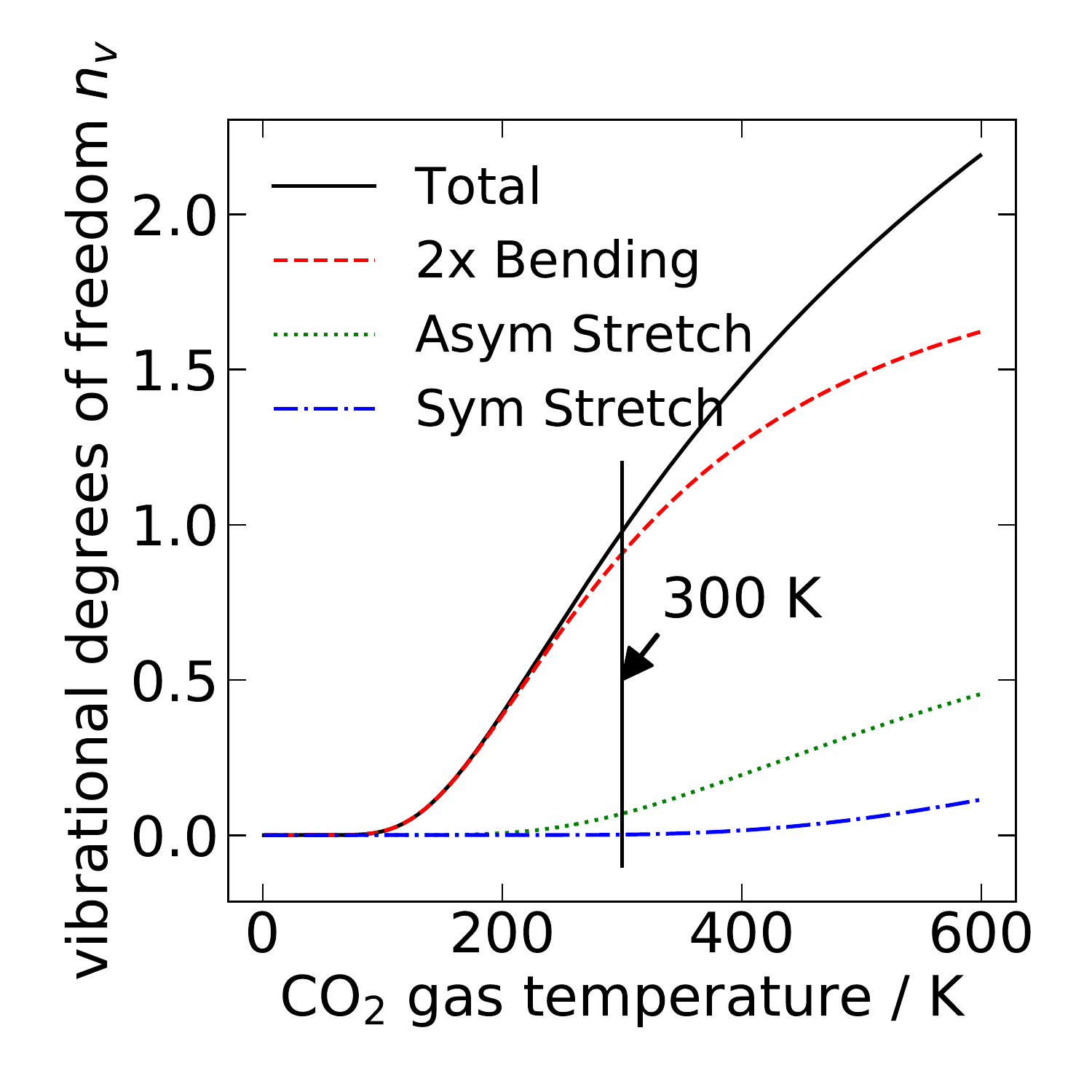}
	\caption{\label{defoffreedom.pdf} %
	Computed temperature dependence of the number of degrees of freedom of the four vibrational modes of carbon dioxide, the bending modes, the asymmetric stretch and the symmetric stretch. See main text and table \ref{tab-vibraitons} for details.
	}
\end{figure}
see ref \cite{atkins} for details, and finally
\begin{equation}
	f = 3 + 2 + 2\sum_v n_v.
\end{equation}
Table \ref{tab-vibraitons} shows the computed $n_v$ values. As expected at room temperature the stretching vibration of nitrogen is not excited and so $n_v \approx 0$ giving overall for nitrogen $f=5$ in good agreement with our measured value and the literature values.

For carbon dioxide the lower energy bending mode is, however, quite excited, that is, it contributes to the heat capacity, whereas the two stretching modes contribute only a fractional amount. Fig. \ref{defoffreedom.pdf} show for CO$_2$ the temperature dependence of the nuber of degrees of freedom for each of its three modes of vibrations and its overall total number of vibrational degrees of freedom. The difference in the stiffness of the three modes is evident in the different computed temperature onsets for each mode. At room-temperature we can see that although the two identical bending modes are excited they are not full excited and so do not make a complete (or integer) contribution to the overall number of vibrational degrees of freedom. Instead for carbon dioxide this analysis predicts $\sum_v n_v = 0.979$ noting that there are two  bending modes with the same energy. This therefore means a predicted value of $f = 6.958$ which is in near agreement with the literature value of $6.83$ and within the uncertainty of our measured value of $6.46 \pm 0.39$. 

A similar analysis could be attempted for the rotational modes, but students may struggle to understand that quantisation, whereas the quantum harmonic oscillator is usually covered in most early years physics courses. 

\subsection{Proposed future work}

As further work the gasses argon and propane are suggested as they lie at the extreme ends of the available $\gamma$ values. Argon is a monotonic gas like helium but its larger mass means it had a much smaller thermal conductivity, more akin to carbon dioxide, and so it maybe possible to prevent, or measure, any heat-transfer to of from the apparatus. Propane has a large number of degrees of freedom and from that it may be able to empirically determine a $\tau$ vs. $f$ relationship to compare with theory. Finally attaching a small vacuum pump to the gas inlet system may reduce the chance of any contamination of the gasses. 

\section{Conclusion}

Through careful early years undergraduate level physics analysis of the experimental apparatus, the collected data and underlying theory, we show that the R\"uchardt experiment is able to measure and differentiate between the adiabatic ratio, and hence degrees of freedom, of several small gasses. We show that a complete description of the work requires mechanical dynamics, electromagnetism, thermodynamics, statistical mechanics and quantum mechanics. This makes it an ideal project to bridge the gap between scripted experiments and more advanced open-ended projects students typically tackle in their final years.

\section{CRediT author statement}
\textbf{Chris Shearwood}: Investigation, Resources, Methodology \textbf{Peter Sloan} Conceptualization, Software, Formal analysis, Data Curation, Writing, Visualization

\section*{References}
\bibliographystyle{unsrt}
\bibliography{export}

\begin{thebibliography}{10}

\bibitem{Ruchardt1929}
E~Rüchardt.
\newblock Eine einfache methode zur bestimmung von c$_p$/c$_v$.
\newblock {\em Phys. Z}, 30:58--59, 1929.

\bibitem{Clark1940}
A~L Clark and L~Katz.
\newblock Resonance method for measuring the ratio of the specific heats of a
  gas, c$_p$/c$_v$: Part i.
\newblock {\em Canadian Journal of Research}, 18a:23--38, 1940.

\bibitem{Hafner1964}
E~M Hafner.
\newblock Refined r\"uchhardt method for $\gamma$.
\newblock {\em American Journal of Physics}, 32:xiii--xiv, 1964.

\bibitem{Smith1979}
D~G Smith.
\newblock Simple c$_p$/c$_v$ resonance apparatus suitable for the physics
  teaching laboratory.
\newblock {\em American Journal of Physics}, 47:593--596, 1979.

\bibitem{Severn2001}
G~D Severn and T~Steffensen.
\newblock A simple extension of r\"uchardt’s method for measuring the ratio
  of specific heats of air using microcomputer-based laboratory sensors.
\newblock {\em American Journal of Physics}, 69:387--389, 2001.

\bibitem{Gruber2004}
Christian Gruber, Séverine Pache, and Annick Lesne.
\newblock On the second law of thermodynamics and the piston problem.
\newblock {\em Journal of Statistical Physics}, 117:739--772, 2004.

\bibitem{Torzo2001}
Giacomo Torzo, Giorgio Delfitto, Barbara Pecori, and Pietro Scatturin.
\newblock A new microcomputer-based laboratory version of the r\"uchardt
  experiment for measuring the ratio $\gamma$=c$_p$/c$_v$ in air.
\newblock {\em American Journal of Physics}, 69:1205--1211, 2001.

\bibitem{Matthias2021}
Matthias Weiszflog and Inga~K Goetz.
\newblock Transforming laboratory experiments for digital teaching: remote
  access laboratories in thermodynamics.
\newblock {\em European Journal of Physics}, 43:015701, 1 2022.

\bibitem{Caccamo_2019}
M~T Caccamo, G~Castorina, F~Catalano, and S~Magazù.
\newblock Rüchardt’s experiment treated by fourier transform.
\newblock {\em European Journal of Physics}, 40:025703, 3 2019.

\bibitem{Bringuier_2015}
E~Bringuier.
\newblock The frictionless damping of a piston in thermodynamics.
\newblock {\em European Journal of Physics}, 36:055024, 9 2015.

\bibitem{CRCthermal}
John~R. Rumble, editor.
\newblock {\em "Physical Constants of Organic Compounds", in CRC Handbook of
  Chemistry and Physics}, volume 103 web version 2022.
\newblock CRC Press/Taylor \& Francis, Boca Raton, FL.

\bibitem{CRCheatcap}
John~R. Rumble, editor.
\newblock {\em "Thermodynamic Properties as a Function of Temperature", in CRC
  Handbook of Chemistry and Physics}.
\newblock CRC Press/Taylor \& Francis, Boca Raton, FL., 103 web version 2022
  edition.

\bibitem{CRCvib1}
John~R. Rumble, editor.
\newblock {\em "Fundamental vibrational frequencies of small molecules", in CRC
  Handbook of Chemistry and Physics}.
\newblock CRC Press/Taylor \& Francis, Boca Raton, FL., 103 web version 2022
  edition.

\bibitem{CRCvib2}
John~R. Rumble, editor.
\newblock {\em "Spectroscopic Constants of Diatomic Molecules", in CRC Handbook
  of Chemistry and Physics}.
\newblock CRC Press/Taylor \& Francis, Boca Raton, FL., 103 web version 2022
  edition.

\bibitem{atkins}
P.~W. Atkins.
\newblock {\em Physical Chemistry}.
\newblock Oxford, 5th edition, 1994.

\end{thebibliography}

\end{document}